# Real-Time Locomotion on Soft Grounds With Dynamic Footprints

Eduardo Alvarado[1]*, Chloé Paliard[2], Damien Rohmer[1] and Marie-Paule Cani[1]

[1]LIX, Ecole Polytechnique/CNRS, Institut Polytechnique de Paris, Palaiseau, France, [2]LTCI, Télécom Paris, Institut Polytechnique de Paris, Palaiseau, France

When we move on snow, sand, or mud, the ground deforms under our feet, immediately affecting our gait. We propose a physically based model for computing such interactions in real time, from only the kinematic motion of a virtual character. The force applied by each foot on the ground during contact is estimated from the weight of the character, its current balance, the foot speed at the time of contact, and the nature of the ground. We rely on a standard stress-strain relationship to compute the dynamic deformation of the soil under this force, where the amount of compression and lateral displacement of material are, respectively, parameterized by the soil's Young modulus and Poisson ratio. The resulting footprint is efficiently applied to the terrain through procedural deformations of refined terrain patches, while the addition of a simple controller on top of a kinematic character enables capturing the effect of ground deformation on the character's gait. As our results show, the resulting footprints greatly improve visual realism, while ground compression results in consistent changes in the character's motion. Readily applicable to any locomotion gait and soft soil material, our real-time model is ideal for enhancing the visual realism of outdoor scenes in video games and virtual reality applications.

## 1 INTRODUCTION

Natural environments present a major challenge when it comes to animating characters in real time, e.g., for video games or virtual-reality applications. Indeed, the diverse conditions that characters may encounter when they move in such environments, from different soil types such as sand or snow to additional flexible elements such as vegetation, give rise to a wide range of possible reactions from the environment and deeply affect characters' motion. Often neglected, this two-way interaction is a major source of non-realism in current applications.

Although considerable efforts have recently been devoted to the use of modern learning methods (deep learning and/or reinforcement learning) to improve character animation, motion usually takes place on rigid grounds. Indeed, the high-dimensional data needed to capture the interaction with various soft grounds would be much more difficult to acquire, and the training of the animation model would be much more intricate in the context of such a dynamic environment.

While standard simulation methods for soft materials could, in theory, be used to solve the two-way interaction between a character and soft grounds in a predictive way, such methods would only be applicable to simulated, physically based characters and may not be usable in real-time applications.

In this work, we address the problem of generating real-time character animation on dynamic, soft grounds, such as sand, snow, soil, or mud. In order to achieve this, two challenges need to be solved: enhancing a standard, kinematic locomotion model to make it adaptive to non-flat terrains and providing a light yet generic ground model, able to react in real time to local interactions while capturing a wide variety of ground behaviors (compressible or not, wet or dry, etc.). Our solution is therefore twofold: we enhance the locomotion model with a global controller that affects a high level, rigid body representation of the character and adds balancing and tilting behaviors to the predefined, forward kinematic animation and we introduce a minimal, physically based model for soft grounds based on a linear strain-stress relationship, associated with a novel way to convert the kinematic state of the character when a foot hits the ground into a plausible interaction force over time. Computing the right interaction force, the function of both foot velocity and the portion of the character weight put there yields the right amount of ground compression and directly leads to a plausible change of subsequent character motion. Moreover, the visually plausible footprints resulting from this interaction are generated without disrupting real-time performance.

We claim the following contributions:

- An enhanced real-time locomotion model that combines a global controller, enabling tilting and balancing behaviors, to an input IK animation, making it adaptable to non-flat terrains (note that we already introduced this model in Paliard et al. (2021).
- A lightweight force model for feet-to-ground contact, driven by the kinematics of the input motion and the nature of the terrain.
- A versatile model based on Hooke's law for ground deformation, parameterized by the Young modulus in compression and the Poisson ratio in lateral material displacement.

We use a real-time implementation in the Unity system to test and validate our method. Our results show that our method gives visually plausible results for a variety of ground types and character gaits and that the interaction forces we compute compare well with real, acquired data for Ground Reaction Forces (GRF).

## 2 RELATED WORK

Our work is related to both real-time animation models for characters, enabling them to adapt to a dynamic environment and the simulation of soft grounds.

### 2.1 Real-Time Characters Interacting With Their Environment

Modeling plausible character motion is a complex problem that couples behavioral intents with physics-based equilibrium constraints. Fully kinematic models were explored in early work (Sun and Metaxas, 2001), leading to real-time adaptation of walking cycles to uneven terrains, at the price of lacking dynamic response to terrain changes. Coupling preset gaits with dynamic events were handled through the help of animation controllers (Yin et al., 2007) that convert differences between the actual pose and the prescribed one as torques, which are then fed into a physically based simulation computing dynamic state changes. Sources of uncertainty can also be included in the control strategy to adapt the motion to unknown situations (Wang et al., 2010). Coupled with reinforcement learning (Peng et al., 2018) and/or adapted reward functions (Kwon et al., 2020), these approaches can model a wide range of plausible complex motions up to acrobatic effects (Won et al., 2020). These motions automatically adapt to their simulated environment at the price of their high computational cost and indirect control and thus cannot be directly applied to loose deforming terrains in standard game engines. In reverse, gathering and processing large pre-recorded motion databases allow for the use of deep-learning methods (Holden et al., 2017), which, once learned, can achieve real-time performance. Still, the need for an extensive motion database would not scale well to the variety of loose environments targeted in this work.

Closer to our goals, simple dynamic models applying forces to the center of mass to control the general balance of a character seen as an inverted pendulum (Kwon and Hodgins, 2010) allow for a good high-level tradeoff between efficiency and plausibility. Mitake et al. (2009) proposed a high-level representation of an oscillating proxy volume, triggering pre-recorded detailed animations (walking, running, falling). This work achieves real-time animation but is restricted to preset animations on flat grounds. Other works modeled the adaptation of a character's gait to more complex environments. Carensac et al. (2015) studied the locomotion of characters partly immersed in a liquid. Bermudez et al. (2018) used drag forces to adapt the character's gait to a fluid environment. Similar to ours, these approaches rely on high-level controllers, but they only take into account the influence of the environment on the character's gait rather than modeling two-way interactions. Moreover, they are restricted to locomotion in a fluid.

### 2.2 Soft Grounds

Terrain deformation is a well-studied area in geoscience and computer graphics (CG) (Galin et al., 2019). At a large scale, terrains are subject to diverse natural phenomena, from the erosion due to sun and water that break and transport ground material (Cordonnier et al., 2017) to the action of wind that piles up sand dunes (Beneš and Toney Roa, 2004) and snow deposit, which creates smooth landscapes (Cordonnier et al., 2018). The effect of moving characters was little studied at such scale, with the exception of skiers leaving tracks and causing avalanches in snow-covered landscapes (Cordonnier et al., 2018) and of the generation of eroded trails due to fauna (Ecormier-Nocca et al., 2021). In both cases, simple textures were used as track models, and none of these works studied the two-way interaction between a deforming ground and a moving creature.

At smaller scales, extensive CG work studied physically based simulation of snow (Stomakhin et al., 2013) and granular material such as dry (Zhu and Bridson, 2005; Ihmsen et al., 2013; Daviet and Bertails-Descoubes, 2016) or wet sand (Lenaerts and Dutré, 2009). Some of these models specifically addressed the two-way coupling of sand simulation with rigid or elastic solids (Narain et al., 2010; Klár et al., 2016). While being able to capture small-scale phenomena, these models are only applicable to a specific type of ground. Moreover, being computationally intensive, they are not applicable to our problem of real time, two-way interactions between characters and soft grounds.

More versatile, approximate models were proposed for real-time applications. Sumner et al. (1999) introduced the first unified model for sand, mud, and snow. They generated foot tracks left by walking characters using a flexible procedural method to move soil material (represented by a hightfield in a 2D terrain grid) away from the intersection area between a foot model and the ground. Onoue and Nishita (2005) introduced a more focused appearance-based approach for granular materials, able to move and accumulate sand on top of colliding bodies. Zeng et al. (2007) further improved this method using Newtonian physics to calculate the deformation. Similar to our work, these two methods cast rays for collision detection between ground and moving bodies and use the impulse experienced by the ground to compute its deformation. However, they did not take the velocity and mass of the colliding bodies into account, thus neglecting the resulting momentum. A more generic, approximate terrain model, composed of a plastic layer on top of an elastic layer, was introduced for simulating ground-vehicle interaction in real time (Zhu et al., 2011). The computation of the interaction force was inspired by Bekker's terramechanics theory (Bekker, 1962), enabling different approximations for snow, grass, sand, and mud. Although our ground model also involves Hooke's low, we make use of a single viscoplastic ground layer, which can be parameterized for different terrain responses. Lastly, Zhu et al. (2019) used a GPU simulation over a 2D grid to achieve sand-vehicle interaction at interactive rates. While their solution is more accurate than ours, it only captures granular materials and provides lower frame rates than our simpler CPU implementation in Unity.

To our best knowledge, the only method that addressed two-way interaction between virtual characters and soft grounds in real time is our recent work (Paliard et al., 2021). An input, kinematic character animation was enhanced with adaptive tilting and dynamic balancing mechanisms thanks to a simple, global controller, leading to automatic gait changes on non-flat terrains. The character poses were driven by inverse kinematics (IK), and ray casting was used to detect collisions between the feet and the ground, enabling the generated footprints to impact the subsequent character's motion. Unfortunately, fully-geometric deformations based on the intersection between the feet geometry and the terrain were used to compute footprints. This method, therefore, failed to account for the dynamics of the interaction, such as the character's quantity of motion, a combination of applied weight and velocity. As a result, a heavy jumping character and a light walking one with similar feet sizes generated the same tracks. While it reuses the same motion controller, our new model includes a physical force model and stress-strain relationships for the soft ground, which, in combination, enables overcoming these strong limitations.

# 3 AN ADAPTIVE MODEL FOR CHARACTER LOCOMOTION

## 3.1 Input Locomotion Model

Our method takes as input an animated character model from a standard game engine, such as a mesh geometry, a rigged skeleton with preset animations, and an IK system applied to feet bones, on top of a simple proxy-geometry and rigid body used for balance control and collision processing. While it allows characters to walk on both flat and sloppy terrains, such models do not allow them to dynamically adapt their gait to varying ground slopes, leading to robotic-looking motions that lack dynamics. Furthermore, they do not provide mechanisms to model any two-way interaction with the environment, restricting the animation to rigid grounds only. Therefore, similar to Mitake et al. (2009), we add to this model a global dynamic behavior, computed by simulating a single rigid body model associated with a bounding capsule. The centroid of the capsule is located at the root joint of the articulated character (the hips), which is assumed to be close to the character's center of mass, and its inertia tensor is precomputed from the character's geometry in rest pose. This simplified physically based model is used for coarse collision detection, enabling the mass, the position of the center of mass, and dimensions of the capsule to have a direct impact on the character's motion. The time evolution of the position and orientation of the physically based capsule is computed from a forward dynamic simulation. It takes as input the velocity of the kinematic animation and a torque. This simulated capsule is then used to define the new orientation and position of the root joint of the character. The limbs corresponding to the feet are finally adjusted in a second step. They are placed using the joint-angles described in the direct kinematics input motion and then adjusted using inverse-kinematics to enforce contact with the ground. The aforementioned elements can be seen as a layered model for character motion control, where each layer acts at a different level of detail. They span from global control for the rigid body that controls the position and orientation of the coarse geometric collider to local control for the IK, whose positions, computed using dynamic ray casting, allow for accurate feet positioning on the ground. To make this model adaptive to non-flat terrains, we enhance the coarser, rigid body model with tilt and swing control terms, as presented next.

## 3.2 A High-Level Controller for Tilting and Swinging Motion

Our global controller computes a torque applied to the rigid body representation of the character over time, modeling the need of the character to self-stabilize on steep terrains. An additional dynamic swinging behavior conveys the effort developed by the character to remain stable and move forward. In contrast to usual controllers, expressed as a difference between current and objective angles for specific limbs, our high-level controller builds on the notion of center of mass, more relevant in our case, by trying to keep its projection along the gravity direction onto the character's support polygon. Because we aim at modeling this tilting motion in the velocity direction, we simplify the representation to a 2D-planar problem, where the plane is defined by the gravity direction, the current character velocity, and is passing through its center of mass. The support polygon is dynamically computed as the segment defined by the projection of the feet positions onto the ground in this plane. The objective projection of the center of mass $p_{com}^{target}$ is thus expressed as the middle of this segment (**Figure 1**-left). The torque magnitude *T* is finally computed as a proportional derivative controller:

$$T = \alpha \left( p_{com}^{target} - p_{com} \right) \cdot u + \beta \dot{\theta}, \quad (1)$$

where $p_{com}$ and $p_{com}^{target}$ are, respectively, the coordinates of the current and the target projections of the center of mass onto the ground, *u* is the unit vector along the terrain slope, and $\dot{\theta}$ is the angular velocity of the rigid body around the axis orthogonal to the plane. The control parameter $\alpha$ is used to tune the freedom of the character to tilt out of its ideal equilibrium position, while $\beta$ sets the amount of dynamic swinging allowed. The values of these parameters can easily be adapted to the nature of the environment, as explained in **Section 6**. As shown in **Figure 1**-middle and right, this new high-level controller automatically adapts to various slopes and character morphology without requiring any parameter tuning. To ensure the accuracy of foot positioning (second, IK layer of the character's model) and avoid any sliding artifact, a small offset is added to the character with the PD controller to match the IK objective of the animation.

## 4 FORCE MODEL FOR FEET-TO-GROUND CONTACT

In this section, we propose a model for the forces that the character applies to the ground when its feet are in contact with it, based on its kinematics and the nature of the ground. The resulting interaction forces over time are used to compute a plausible ground deformation as described in the second part of **Section 5**. The overview of our global physics-inspired terrain deformation process is displayed in **Figure 2**.

### 4.1 Static Forces Mode and Weight Ratio

When a foot of the character is in contact with the terrain, it exerts a force on it that we denote $\vec{F}_{foot}^{left}$ and $\vec{F}_{foot}^{right}$ for the respective left and right foot, or denoted simply $\vec{F}_{foot}$ for a single foot when its left/right placement is not important. This force is at the origin of the deformation of the terrain, and we therefore aim to estimate as long as the foot is considered to be in contact with the ground. $\vec{F}_{foot}$ is closely related to the notion of the so-called Ground Reaction Force $\vec{F}_{GR}$, which is the reaction force that the terrain exert on the foot and we have at any time of the contact:

$$\vec{F}_{GR} = -\vec{F}_{foot}. \quad (2)$$

While we will consider later the geometrical contact surface between the foot and the terrain to compute the deformation of the ground, we assume at this stage that the contact forces are applied at the foot's joint position, supposed to be a fixed point relative to the character.

When the character is static, $\vec{F}_{foot}$ only depends on its weight $\vec{F}_{weight}$ and on the current character balance. Let *m* be the mass of the character and $\vec{g}$ the gravity constant. We can write

$$\vec{F}_{foot}^{left} + \vec{F}_{foot}^{right} = \vec{F}_{weight}^{left} + \vec{F}_{weight}^{right} = m\vec{g}. \quad (3)$$

If only a single foot is in contact with the ground, then the full weight of the character is applied through it, while the other foot has a null contribution with respect to the ground. When the two feet are in contact, the weight force is distributed through the two contact points, and we introduce the weight ratios $\alpha^{left}$ and $\alpha^{right}$ such that

$$\vec{F}_{weight}^{left/right} = \alpha^{left/right} \, m \, \vec{g} \quad (4)$$
$$\text{with } \alpha^{left/right} \in [0, 1], \text{ and } \alpha^{left} + \alpha^{right} = 1.$$

Assuming that the mass of the character is evenly distributed in the 2D plane defined by the contact points of the feet $p_{left}$, $p_{right}$ and its center of mass, the weight ratio represents the relative position of $p_{com}$—the projection of the center of mass on the segment $[p_{left}, p_{right}]$—leading to

$$\alpha^{right} = \frac{\left( p_{com} - p_{left} \right) \cdot \left( p_{right} - p_{left} \right)}{\| p_{com} - p_{left} \| \, \| p_{right} - p_{left} \|}. \quad (5)$$

This weight ratio can be intuitively seen as a way to split the contribution of the character weight between a left and right side that adapts to its posture (Garcia, 2012) and is therefore a value that is re-computed at each frame of the animation.

### 4.2 Dynamic Force Model During the Contact

When the character is in walking motion, its global change of momentum is expressed by forces exerted on the ground and applied at the position of the feet. Let us call $\vec{F}_{momentum}$ the additional force due to the motion such that

$$\vec{F}_{foot}^{left/right} = \vec{F}_{weight}^{left/right} + \vec{F}_{momentum}^{left/right} \quad (6)$$

Our idea is to estimate plausible value for $\vec{F}_{momentum}$ from the kinematics inputs. To this end, we will make the assumption that, during the gait phase where the foot moves down and touches the ground, the corresponding left/right side of the character—as induced by our definition of $\alpha^{left/right}$—acts like a rigid body falling down. Therefore, we neglect the influence of the internal muscle that would lead to local adaptation of the articulations during the impact and assume that the change of momentum of each side of the character is fully supported by the ground.

When the character's foot touches the ground at a time $t_0$, the IK-based system considers its position to be fixed, and its motion stops therefore suddenly. This change of velocity can be represented by an impulse $\vec{J}$ of the feet on the ground. Considering our previous assumption that a whole part of the body falls rigidly, we express this impulse as

$$\vec{J} = \alpha\, m\, \vec{v}(t_0 - \Delta t), \qquad (7)$$

where $\vec{v}(t_0 - \Delta t)$ is the velocity of the foot evaluated at the frame preceding its impact on the ground and $\Delta t$ is the time step of the real-time game engine. A naive approach could consist in computing a short momentum force $\vec{J}/\Delta t$ during the time step $\Delta t$, this would, however, lead to a very large force burst during a single time step of the animation, which may not accurately represent the slower motion happening on soft ground as the foot velocity obtained from the forward kinematics and IK constraints does not represent this specific foot-to-soft-ground interaction. Our objective is thus to model a simple yet reasonable continuous momentum force expression during contact with such inaccurate kinematics input. To this end, we propose to introduce an external parameter called the characteristic time $\tau$, possibly greater than $\Delta t$, that represents the typical time needed for the character to be fully stopped by a given type of terrain. In a first approximation, we will assume, for a standard range of weight and velocity compatible with human locomotion, the value $\tau$ as a constant that only depends on the type of terrain. Thanks to this parameter, we can model the momentum force as the vector:

$$\vec{F}_{momentum} = \vec{J}/\tau = \alpha\, m\, \frac{\vec{v}(t - \Delta t)}{\tau}, \qquad (8)$$

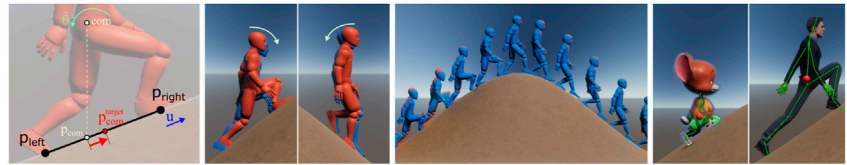

**FIGURE 1** | Adaptive character locomotion on non-flat terrains. Left: the torque magnitude depends on the distance between the middle of the support polygon $p_{com}^{target}$ and the projection of the character's center of mass $p_{com}$ on the terrain, as well as on the current angular velocity $\dot{\theta}$. Middle: comparison between characters without (red) and with (blue) our torque-based controller. The blue character exhibits an increased swinging motion, which dynamically tilts it forward or backward, depending on the slope. Right: the controller is robust to changes of character morphology as the oscillating motion automatically adapts when the center of mass (red sphere) is positioned closer to or further from the ground.

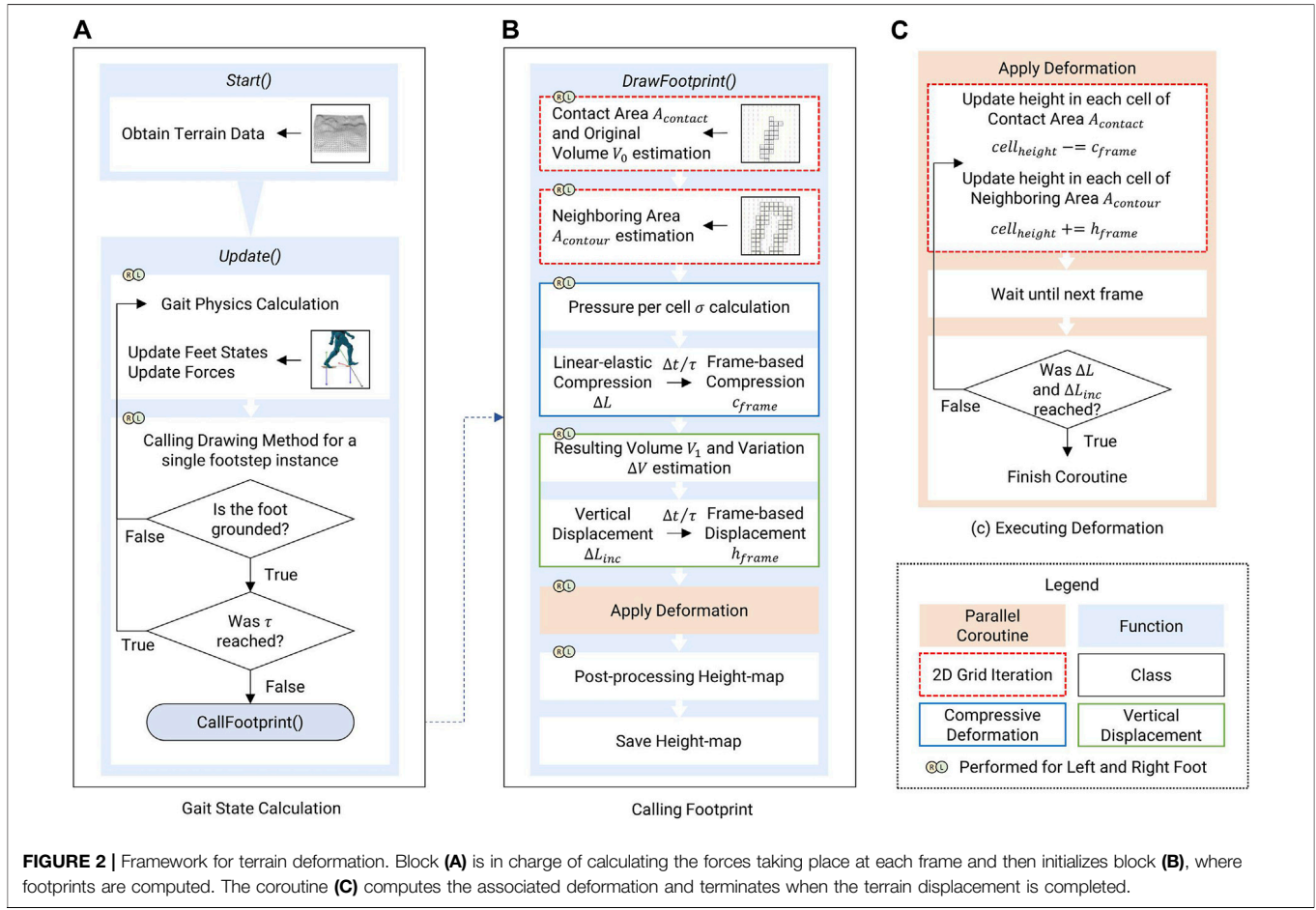

**FIGURE 2** | Framework for terrain deformation. Block **(A)** is in charge of calculating the forces taking place at each frame and then initializes block **(B)**, where footprints are computed. The coroutine **(C)** computes the associated deformation and terminates when the terrain displacement is completed.

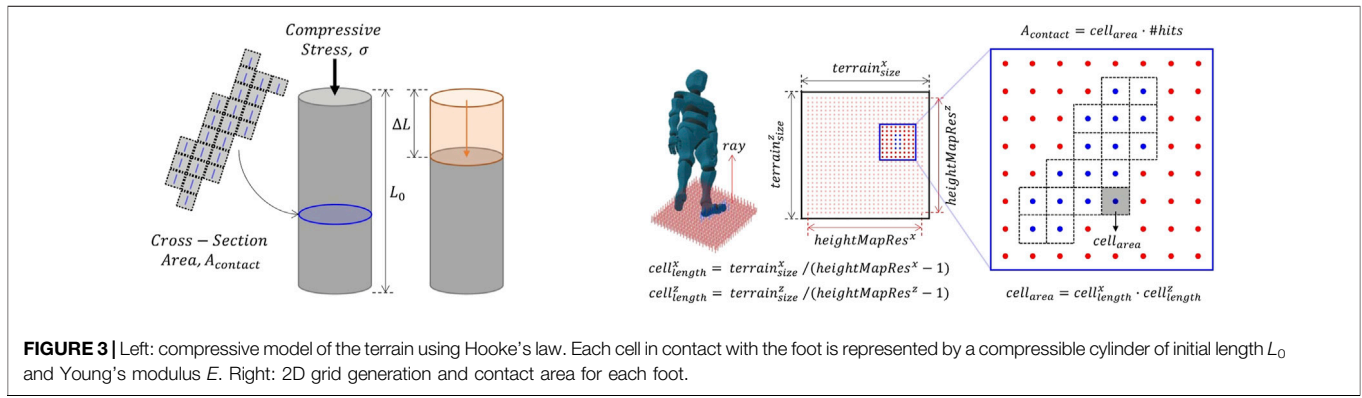

**FIGURE 3** | Left: compressive model of the terrain using Hooke's law. Each cell in contact with the foot is represented by a compressible cylinder of initial length $L_0$ and Young's modulus $E$. Right: 2D grid generation and contact area for each foot.

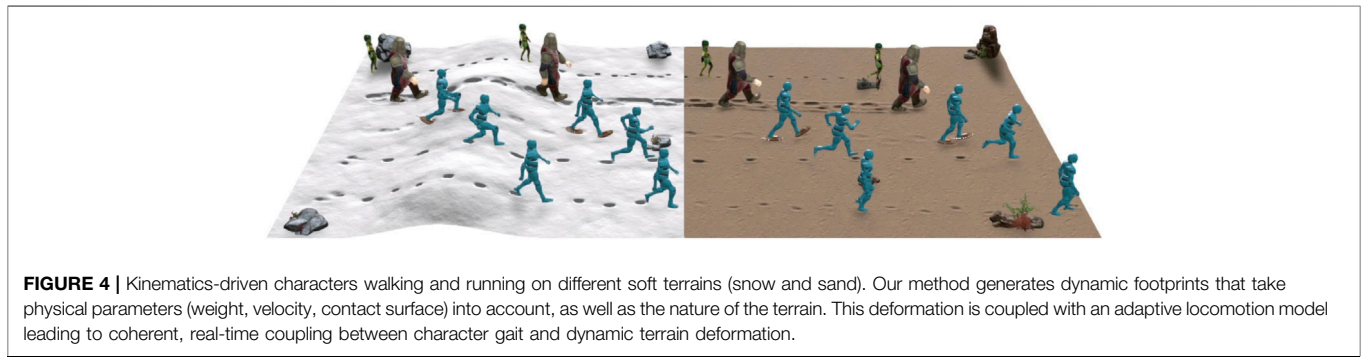

**FIGURE 4** | Kinematics-driven characters walking and running on different soft terrains (snow and sand). Our method generates dynamic footprints that take physical parameters (weight, velocity, contact surface) into account, as well as the nature of the terrain. This deformation is coupled with an adaptive locomotion model leading to coherent, real-time coupling between character gait and dynamic terrain deformation.

applied from the colliding foot to the ground during the time interval $t \in [t_0, t_0 + \tau]$, where $t$ is the current frame time. Note that this formulation allows scaling the magnitude of $\vec{F}_{momentum}$ independently from the engine frame rate and using instead the parameter $\tau$ that relates to the type of terrain. Therefore, a given forward kinematics motion provided as input can be associated with different forces, such as a large magnitude of momentum force on a hard terrain with a small $\tau$ value and a small force magnitude with a long effect on soft terrain with large $\tau$. This dependence of $\vec{F}_{momentum}$ magnitude and duration time to the type of ground is then used to modify the way the terrain is modified as explained in the following section.

## 5 REAL-TIME TERRAIN DEFORMATION

Let us now describe the second part of our two-way interaction model, namely terrain deformation, which in turn will modify the character's gait.

### 5.1 Plastic Model for Terrain Compression

Our first objective is to convert the time-varying force $F_{foot}$ into a dynamic compression behavior that will take place under the character's foot. Our model is based on the following two assumptions. First, we suppose that only the upper layer of the ground is loose enough to be deformed by the action of the foot. This loose upper layer of depth $L_0$, typically a few centimeters, depends on the material, and we suppose that the deeper part below is fully rigid. Second, we assume that this loose layer exhibits a viscoplastic deformation following Hooke's law under compression. Once the foot stress is released, we assume a pure plastic behavior, and the footprint remains fully carved in the terrain. As illustrated in **Figure 3**-left, each grid cell of the terrain is seen as a compressive rod, with relative deformation characterized by its Cauchy strain $\epsilon = \Delta L/L_0$, which is linearly related to the stress $\sigma$ exerted by the foot through its Young modulus of elasticity $E$ and $\sigma = E \epsilon$. The plastic behavior of the material is enforced in keeping track of a total accumulated compression $c_{acc} \in [0, L_0]$ for each cell of the terrain grid and constraining it to only increase from one frame to the other.

To compute the stress $\sigma$ applied to the ground, we consider the orthogonal component of the force $F_{foot\perp}$ with respect to the ground normal $\vec{n}$ such that $F_{foot\perp} = \vec{F}_{foot} \cdot \vec{n}$ and assume that this component is evenly spread on the contact area $A_{contact}$ between the foot and the ground. This leads to the following stress

$$\sigma = F_{foot\perp} / A_{contact}. \quad (9)$$

The contact area $A_{contact}$ must be efficiently computed as it varies dynamically with the motion of the character and the deformation of the terrain. We first set a local window on the terrain grid set as a bounding square around the initial foot position at the contact time (**Figure 3**-right). Then, at each frame, we use ray casting on each cell of the bounding square to estimate the total area where the foot collides with the terrain. Considering that each terrain cell has a unit area $cell_{area}$ and the ray cast detects

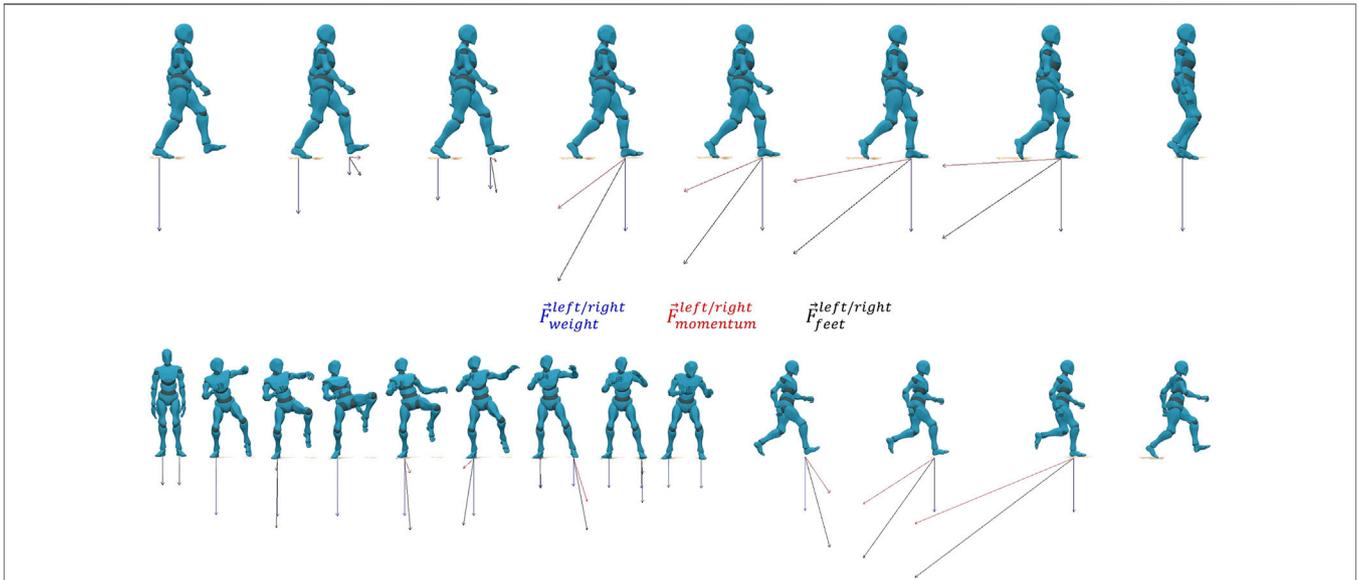

**FIGURE 5** | Display of the force vectors $\vec{F}_{weight}$, $\vec{F}_{momentum}$, $\vec{F}_{foot}$ for each foot in contact with the terrain. The distribution of force automatically adapts to the left and right foot depending on the posture of the character (top and bottom left). Higher velocity during a running motion increases the magnitude of the force (bottom right).

#hits cells where the foot collides with the terrain, we have $A_{contact} = \#hits \cdot cell_{area}$.

Based on our compressible rod model, we can therefore compute a target deformation magnitude:

$$\Delta L = \sigma \frac{L_0}{E} = \frac{\vec{F}_{foot} \cdot \vec{n}}{A_{contact}} \frac{L_0}{E}. \quad (10)$$

While this target deformation is dynamically updated at each frame, we cannot apply it instantaneously to the terrain as it does not take into account its viscoplastic behavior. The latter indeed limits its deformation rate along the characteristic time $\tau$. Instead, we apply on each cell the limited per-frame compression $c_{frame}$ given by

$$c_{frame} = \begin{vmatrix} \Delta t/\tau \ \Delta L, & \text{if } c_{acc} < \Delta L \leq L_0 \\ 0 & \text{otherwise} \end{vmatrix} \quad (11)$$

with $c_{acc}$ adding the new value of $c_{frame}$ at each new frame.

## 5.2 Terrain Vertical Accumulation Using Poisson's Ratio

While some terrain types such as snow can be highly compressible, others such as dry sand are not. In this case, the volume carved under the foot needs to be moved transversely and accumulated to form a bump around the foot position. We model this behavior by assuming that a given terrain can be homogeneously characterized by Poisson's ratio $\nu$ with small deformation such that

$$\frac{V_1 - V_0}{V_0} = (1 - 2\nu) \frac{\Delta L}{L_0}, \quad (12)$$

with $V_0 = A_{contact} L_0$ being the volume before deformation and $V_1$ the final volume corresponding to this material after deformation. Let us call $V_1^c = A_{contact} (L_0 - \Delta L)$ the volume corresponding to a purely compressible material. Then, the additional amount of volume characterized by Poisson's ratio trying to extend in the orthogonal direction of the compression is $\Delta V = V_1 - V_1^c$. We will now assume that this additional amount of material is constrained to accumulate only at the top of the terrain and around the foot, thus creating a local bump.

Let us call $A_{contour}$ the area defined as the exterior neighborhood of $A_{contact}$. The diameter of $A_{contour}$ is denoted by $r_{contour}$, which represents how far the accumulated material may spread around the feet on a given type of ground. $A_{contour}$ is represented on the discrete terrain grid as a finite number of #neighbors cells. Assuming a uniform distribution of material over $A_{contour}$, each cell of this area should follow a target height increase of

$$\Delta L_{inc} = \frac{\Delta V}{cell_{area} \ \#neighbors}. \quad (13)$$

Similar to the compression, this target increase of material is also temporally spread along the characteristic time $\tau$ and we therefore apply at each frame an actual height increase $h_{frame}$ of

$$h_{frame} = \begin{vmatrix} \Delta t/\tau \ \Delta L_{inc}, & \text{if } h_{acc} < \Delta L_{inc} \\ 0 & \text{otherwise} \end{vmatrix} \quad (14)$$

with $h_{acc}$ begin the accumulated height on the cell of $A_{contour}$.

To preserve a smooth appearance of the bumps, we display a filtered version of the accumulated heightfield using standard Gaussian blur. In order to avoid accumulating volume errors, we keep in memory and use the uniform accumulated height for computations, while the smoothed version is only used for display.

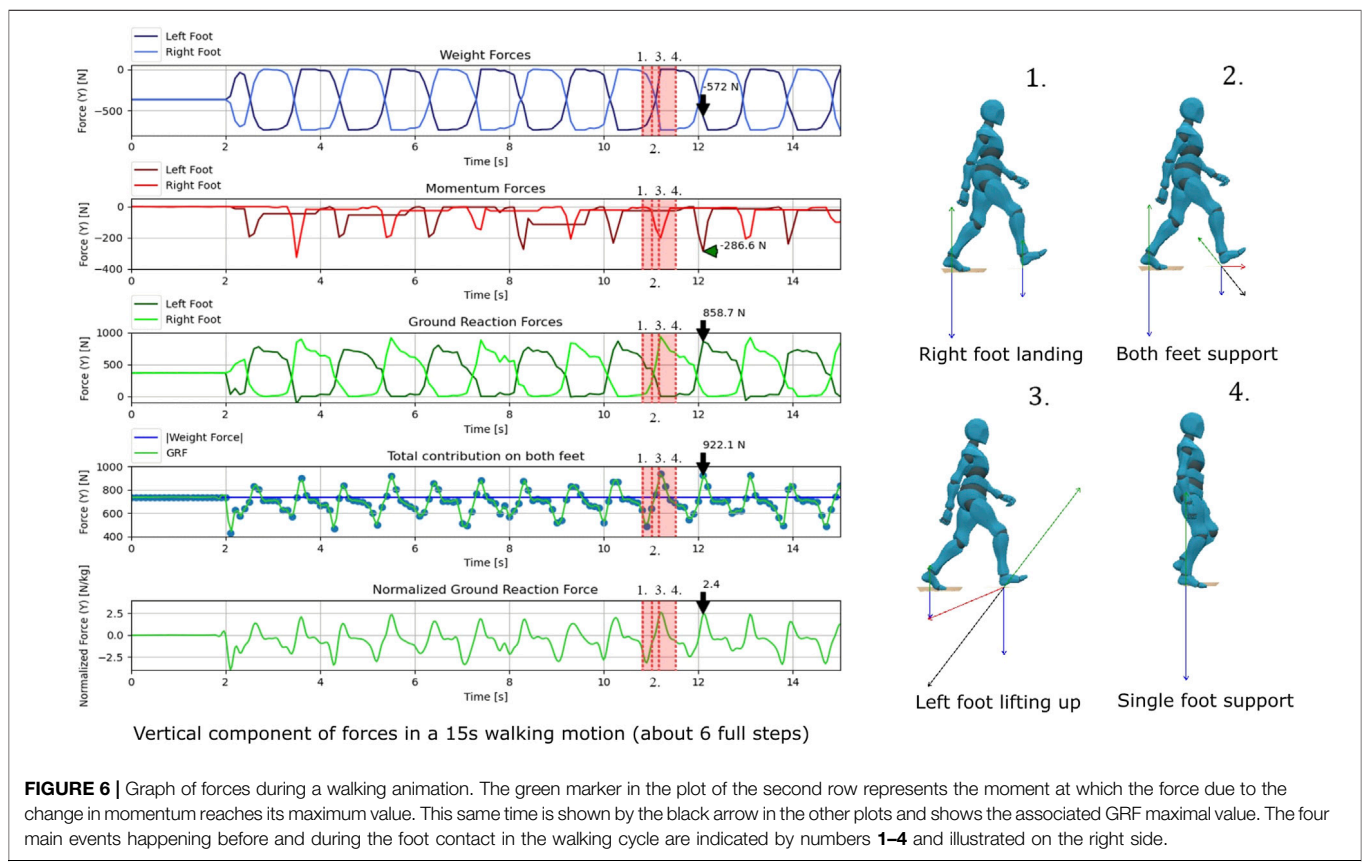

**FIGURE 6** | Graph of forces during a walking animation. The green marker in the plot of the second row represents the moment at which the force due to the change in momentum reaches its maximum value. This same time is shown by the black arrow in the other plots and shows the associated GRF maximal value. The four main events happening before and during the foot contact in the walking cycle are indicated by numbers **1–4** and illustrated on the right side.

## 6 RESULTS AND APPLICATION

We implemented our method as an interactive prototype in Unity (**Figure 4**), where the character can be controlled using the keyboard or a game pad. In the last case, the character's motion will interpolate between walking, running, and jumping animation, based on the intensity with which we move the joystick. The character is placed on different preset terrain types, such as snow, dry sand, and mud. Each of these sample grounds has a size of 10, ×, 10 m with a maximum vertical difference of 10 m between the lowest and highest point. The height maps are initially sampled using a grid of $256^2$ points. We use a bounding square of size 40 × 40 cm to subdivide the terrain grid around each foot impact center. To estimate the neighboring area where the vertical deformation around the impact is applied, we define $r_{contour} = 4\ cm$. The loose upper layer length in the linear-elastic model is described by $L_0 = 30$ cm. The default character has a total mass of 77.5 kg and a height of 1.8 m. The feet are defined by abstract box colliders of a surface area of 10 cm$^2$. The locomotion controller was set with values $\alpha = 30N$ and $\beta = 6Nm$. Our deformation implementation in Unity makes use of coroutines which allow computing the terrain modification using a non-blocking multitask process. The process associated with the deformation (the orange blocks in **Figure 2**) runs parallel with the character IK and our model of controller and is automatically synchronized for coherent display. The entire method—character locomotion and terrain deformation—is fully coded in high-level C# scripts but still runs in real time at about 65 fps on a standard laptop (CPU: Intel Core i7, eight cores, 3.10 GHz) for a resolution of 4 cm (grid of $256^2$ shown in our illustrations) and 30 fps for a resolution of 2 cm per cell.

### 6.1 Validation of Ground Reaction Forces

We first show that our model of force robustly adapts to various situations, including walking motion and running, or equilibrium situation where the character stands alternatively with one or two feet in contact with the ground. **Figure 5** displays the vectors corresponding to $\vec{F}_{weight}$, $\vec{F}_{momentum}$, and $\vec{F}_{foot} = -\vec{F}_{GR}$ for the left and right foot at different times of character animations. Figure 5 illustrates the impact of the foot velocity prior to contact and the distribution of forces between the left and right foot.

We then display and compare the values of our estimated Ground Reaction Forces on a standard walking motion with respect to real measurements from the literature (Ren et al., 2008). For this experiment, we considered a relatively hard terrain defined by a small characteristic time $\tau = 0.05s$ and Young's modulus of 2 MPa for small terrain deformation. We set the subject to have similar weight and height with respect to the average real subjects and set the walking speed to 1.65 m/s. The corresponding graph of the forces is reported in **Figure 6**. The first row shows how each foot supports the proportional part of the weight depending on the gait phase. When the right foot starts to hit the terrain 1), it prompts the ground to respond with an additional force due to the variation in momentum, as shown in the second row. This affects, in return, the Ground Reaction Forces by inducing a peak in that exact moment, as illustrated in the plot of the third row, and continues during the short time when both feet are grounded 2). The force exerted due to the impact mitigates when the character begins to place most of its weight on the right foot 3), being the instant at which the GRFs stabilizes with the force caused by the character's weight. The switch between the forces is complete when the right foot supports the entire weight of our model 4).

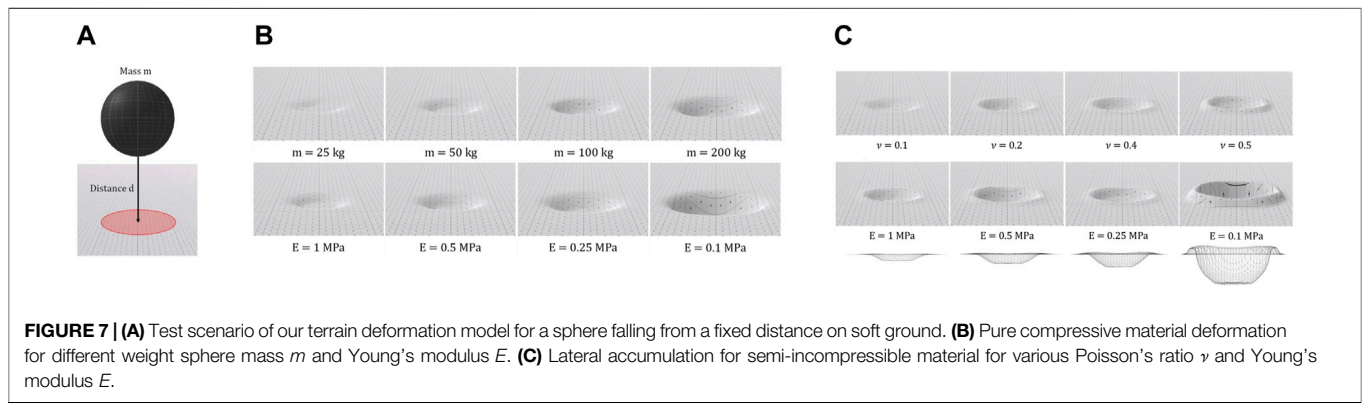

**FIGURE 7 | (A)** Test scenario of our terrain deformation model for a sphere falling from a fixed distance on soft ground. **(B)** Pure compressive material deformation for different weight sphere mass $m$ and Young's modulus $E$. **(C)** Lateral accumulation for semi-incompressible material for various Poisson's ratio $\nu$ and Young's modulus $E$.

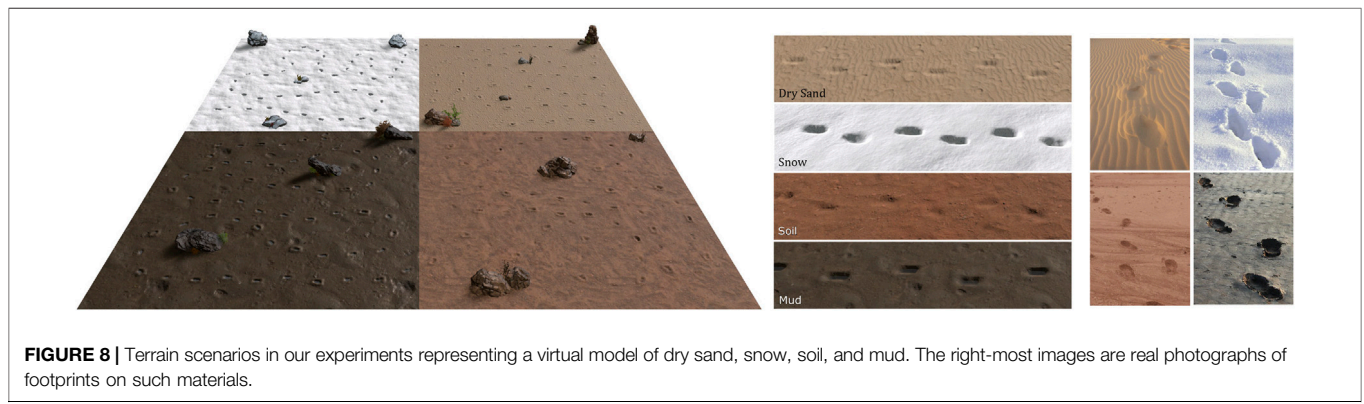

**FIGURE 8 |** Terrain scenarios in our experiments representing a virtual model of dry sand, snow, soil, and mud. The right-most images are real photographs of footprints on such materials.

**TABLE 1 |** Parameters used to represent the different type of terrains.

| Parameter | Snow | Dry sand | Mud | Soil |
|---|---|---|---|---|
| Young's modulus $E$ (MPa) | [0.25, 0.5] | [1.5, 3] | [0.5, 0.75] | [1, 1.5] |
| Characteristic time $\tau$ (s) | 0.2 | 0.05 | 0.15 | 0.05 |
| Poisson's ratio $\nu$ | 0.05 | 0.5 | 0.35 | 0.35 |
| Gaussian filter variance (cm) | 0.5 | 1 | 1 | 0.5 |

These graphs show the very similar shape and range of magnitude with the real measurements obtained using force plates and optical capture system reported by Ren et al. (2008) and Shahabpoor and Pavic (2017). In particular, the main steps of the walking cycle (1, 2, 3, 4) are depicted on the graph of forces, and the total vertical component of Ground Reaction Force normalized by the character weight varies has a maximum pic intensity in the range of 2.5–3.

## 6.2 Test Case of Terrain Deformation

We present in **Figure 7** a test scenario of our terrain deformation for the simple case of a sphere of mass $m$ falling from a distance $d = 2.5$ m. When not explicitly modified, the default parameters are $m = 50$kg, $E = 1$MPa, and $\tau = 0.2$s. **Figure 7B** shows the increase in the terrain hole for a purely compressive material with varying mass and Young's modulus. **Figure 7C** shows the full model of a semi-compressible material, with the effect of increasing the Poisson ratio (constant hole depth, but increased amount of accumulated material) and decreasing the Young modulus (deeper deformation leading to a larger amount of moved material).

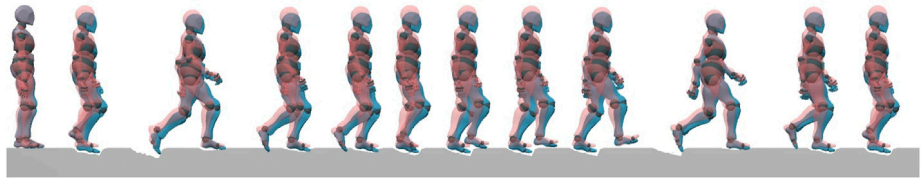

**FIGURE 9** | Dynamic adaptation of the character's walking gait to the change of the terrain. Red: default character walking on flat ground. Blue: our result, with a character, locally adapting its posture to the evolving terrain.

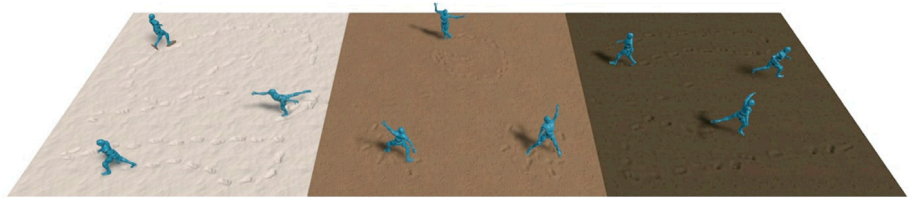

**FIGURE 10** | Various kinematics motion inputs lead to plausible terrain footprints.

## 6.3 Application to Various Ground Materials

We are able to model different types of ground material, as illustrated in **Figure 8**. The latter were set coherent with real photographs of footprints, using the parameters reported in **Table 1**. Young's modulus values were given the same order of magnitude as the ones reported by Gerling et al. (2017), while the characteristic time, Poisson's ratio, and Gaussian filter were set based on phenomenological tests.

Thanks to the automatic adaptation of the character locomotion controller to the feet position on the terrain, our system handles two-way, dynamic interaction between the character gait and the terrain deformation. Such an effect on the character motion is illustrated in **Figure 9**, comparing the default walking behavior on a flat terrain (red character) to the one considering the change of terrain (blue character).

**Figure 4** illustrates a typical locomotion scenario with a variety of behaviors, terrain geometry on snow, and sand model. The closest character follows a standard walking motion, and we can note the change of footprint depth when moving from snow to sand. The second character is running. Due to its velocity, the footprints remain deep on both types of ground. Note that our locomotion system automatically adapts its legs' position on the small hill. The third character wears snowshoes that increase the surface of its feet, thus leading to shallow footprints. The increased size of the feet leads to the adaptation of the gait when a collision with the terrain is detected. The fourth character is heavy, which leads to deep footprints. The character also has large feet volume and small legs compared to its size, which coupled to the deep prints, leading to an almost constant collision between the feet and the ground, which turns out to generate trails. The last walking character has a smaller weight and a tiny foot surface, which leads salient, individual footprints. We further encourage reviewers to watch the full quality video accompanying this submission at the following URL[1].

Finally, thanks to its simplicity, our method is able to robustly extend to various types of kinematics input beyond standard walks and can generate plausible footprints along the feet trajectory, as illustrated in **Figure 10**.

## 7 DISCUSSIONS AND CONCLUSION

We have proposed a method able to convert, in real time, a simple kinematic locomotion model into a dynamic animation with two-way interactions between the character and soft grounds. Our physically based interaction forces drive the generation of local footprints, controlled using standard values parameters for the ground material such as its Young's modulus and Poisson's ratio. Given a user-defined terrain type, our method is able to compute coherent forces during foot contact, despite considering a raw kinematics input for the character model. The geometric deformation of the terrain is itself impacting the walking gait, thus leading to a bilateral interaction between the soft terrain and the motion of the character. The main advantage of this approach is to be lightweight enough to run interactively in a game engine like Unity. Beyond interactive video-game applications, such a system could be useful in virtual reality applications, with a VR setup where the user could see in real time its own footprints. Our force model could also be coupled with other physically based systems to allow interaction with other objects in the scene.

---

[1]https://drive.google.com/drive/folders/1Bke_ NmZLUEU0fKLgWMHt7j5128Myqvz_?usp=sharing.

We made a series of approximations in our force model to achieve efficient computations, leading to several limitations. In order to estimate the interaction forces between the character and the ground from kinematic inputs, we only considered the change of velocity of the foot and assumed that the character behaves like a rigid body falling down, thus neglecting any internal muscle influence that the character might use for dynamically adapting its motion during the collision of its feet with the terrain. While this assumption led to reasonable results for walking characters, it could not capture the change of pressure exerted on the ground during foot kick, or dancing-like motion, for instance, where body parts in the air have rapid variations of momentum. To accurately handle such cases, the change of momentum of each body part could be taken into account in a global system to retrieve all the forces and torque at all joints. Furthermore, the use of a fully controlled-based model together with an optimization strategy for each joint could bring an advantage, not only to reproduce more suitable deformations but also to better recreate the impact of the terrain on the character's movement. This might be helpful in different situations, such as decreasing the character's speed on terrains where his feet sink to reduce effort and maintain a constant energy consumption or extending the arms and bending the knees for balance and stability.

In addition, our physical model for grounds, which makes use of a global property for the deformable material, is also an approximation of how granular material behaves. In particular, assuming a constant deformation time $\tau$ per material is only valid within a certain range of mass and velocity of the character. Making $\tau$ dependent on velocity in the case of a jumping character, for instance, would allow modeling a fast ground deformation immediately after the impact and slowing down as the velocity is absorbed. The vertical accumulation also only approximates the final volume of displaced material as being spatially filtered by a Gaussian. As an extension, we would like to integrate the direction and magnitude of the velocity to impact the resulting bump and integrate a local terrain tessellation so that our method could be applied to arbitrary terrain dimensions.

Finally, we would like to adapt the system to synthesize the impact of not only bipedal characters but also quadrupeds. To this end, we need to adjust the system so that the total mass of the character is correctly distributed based on a variable number of contact points with the ground. Having such low-level deformations on the terrain could help us better recreate the long-term impact of virtual animals on natural environments, as an extension of Ecormier-Nocca et al. (2021).

## DATA AVAILABILITY STATEMENT

The datasets presented in this study can be found in online repositories. The names of the repository/repositories and accession number(s) can be found in the following link: https://github.com/edualvarado/unity-footprints.

## AUTHOR CONTRIBUTIONS

EA, DR, and M-PC contributed to the conception and design of the research project. EA is the main contributor and developed most of the technical part of the project, conducted the experiments shown in the results section, and programmed the code available in the **Supplementary Material**. CP was the main contributor of the associated short paper "Soft Walks: Real-Time, Two-Ways Interactions between a Character and Loose Grounds, EG2021" at the starting point of this research, and developed the high-level controller described in Sec. *A High-Level Controller for Tilting and Swinging Motion*. All authors contributed to the manuscript redaction and revisions.

## FUNDING

This work received funding from the European Union's Horizon 2020 Research and Innovation Programme under the Marie Skłodowska-Curie Grant Agreement n. 860768 (CLIPE project).

## SUPPLEMENTARY MATERIAL

The Supplementary Material for this article can be found online at: https://www.frontiersin.org/articles/10.3389/frvir.2022.801856/full#supplementary-material